\documentclass[paper]{JHEP3}
\pdfoutput=1
\usepackage[pdftex]{graphicx}
\usepackage{amsfonts,amsmath}

\title{Observational constraints on the $\Lambda$LTB model}

\author{
Valerio Marra\footnote{valerio.marra@jyu.fi} \phantom{ }and
Mikko P\"a\"akk\"onen\footnote{mikko.u.paakkonen@jyu.fi}\\
Department of Physics, PL 35 (YFL),\\
FI-40014 University of Jyv\"askyl\"a, Finland,\\
and\\
Helsinki Institute of Physics, PL 64,\\
FI-00014 University of Helsinki, Finland.
}

\abstract{
We directly compare the concordance $\Lambda$CDM model to the inhomogeneous matter-only alternative represented by LTB void models.
To achieve a ``democratic'' confrontation we explore $\Lambda$LTB models with non-vanishing cosmological constant and perform a global likelihood analysis in the parameter space of cosmological constant and void radius.
In our analysis we carefully consider SNe, Hubble constant, CMB and BAO measurements, marginalizing over spectral index, age of the universe and background curvature.
We find that the $\Lambda$CDM model is not the only possibility compatible with the observations, and that a matter-only void model is a viable alternative to the concordance model only if the BAO constraints are relaxed.
Moreover, we will show that the areas of the parameter space which give a good fit to the observations are always disconnected with the result that a small local void does not significantly affect the parameter extraction for $\Lambda$CDM models.
}

\preprint{arXiv:1009.4193}

\keywords{dark energy theory, supernova type Ia - standard candles, cosmological parameters from CMBR, baryon acoustic oscillations}

\begin{document}

\section{Introduction} \label{introduction}

In the past 15 years it has been extensively studied (see for example \cite{Moffat:1994qy,Mustapha:1998jb, Tomita:1999qn, Celerier:1999hp, Iguchi:2001sq, Alnes:2005rw, Chung:2006xh, Enqvist:2006cg, Tanimoto:2007dq, Alexander:2007xx, GarciaBellido:2008nz, Clifton:2008hv, Yoo:2008su, Zibin:2008vk, February:2009pv, Sollerman:2009yu, Kolb:2009hn, Dunsby:2010ts, Yoo:2010qy, Biswas:2010xm, Clarkson:2010ej, Moss:2010jx}) how an observer inside a matter-only spherical void expanding faster than the background sees apparent acceleration.
This effect is easy to understand: our cosmological observables are confined to the light cone and hence temporal changes can be associated with spatial changes along photon geodesics. In the present case, ``faster expansion now than before" is simply replaced by ``faster expansion here than there".  This is why a void model can mimic the effect of a cosmological constant if it extends to the point in space/time where the dark energy becomes subdominant.
A typical scenario that can mimic the late-time acceleration of the concordance $\Lambda$CDM model consists of a deep void extending for 1-3~Gpc.

Void models feature the heavy fine tuning of the observer's position which, to be consistent with the CMB dipole, has to be very close to the void center (few percents \cite{Kodama:2010gr,Foreman:2010uj}, se also \cite{Quartin:2009xr}), thus leading to a violation of the Copernican principle.
However, it is not clear how to regard this fine tuning as compared, for example, to the fine tuning of the cosmological constant or of other dark energy sources, and a pragmatic approach is just to rely on the observational data in order to find the most satisfying model.

In the present work we adopt the latter pragmatic approach and to ``democratically'' find the best-fit model we study void models with cosmological constant.
More specifically we will consider the so-called $\Lambda$LTB models, which are spherically symmetric dust Lema\^{i}tre-Tolman-Bondi (LTB) models with non-vanishing cosmological constant \cite{Lemaitre:1933gd,Tolman:1934za,Bondi:1947av}.
We will run likelihood analyses over two parameters: one (the density parameter $\Omega_{\Lambda}$) pertinent to the $\Lambda$CDM model and the other (the void radius $r_{0}$) pertinent to the LTB model.\footnote{
Also the void depth strongly characterizes a void model and we will keep it fixed to the value required by a good fit to the SNe data, see Section \ref{results}.}
Our parameter space will therefore contain both the $\Lambda$CDM model and the matter-only LTB model as delimiting axes, allowing a direct confrontation between the two opposite alternatives and overcoming some of the ambiguities in comparing the two models.
We remind that the LTB model is specified not by free parameters, but by free functions and therefore the number of degrees of freedom to be adopted in computing the reduced~$\chi^{2}$, which is usually used in model ranking, is necessarily a somewhat subjective choice.
Our analysis will also show if the data favor a mixed scenario of a void together with a cosmological constant, which is a possibility not yet fully explored.

It is worth pointing out that the study of void models represents just a small (even if admittedly the most successful) part of the research devoted to understand how large-scale inhomogeneities affect the observations, a phenomenon collectively referred to as backreaction (see \cite{Kolb:2009rp}).
Weak backreaction studies focus on the observational properties of the universe and include, besides the void models here examined, also swiss-cheese \cite{Kantowski:1969, Marra:2007pm, Marra:2007gc, Biswas:2007gi, Brouzakis:2007zi, Vanderveld:2008vi, Bolejko:2008xh, Valkenburg:2009iw}, onion \cite{Biswas:2006ub, Marra:2008sy} and meatball \cite{Kainulainen:2009sx, Kainulainen:2009dw, Clifton:2009jw, Amendola:2010ub} models.
Strong backreaction studies, on the other hand, address the question of how the cosmological background reacts to the nonlinear structure formation, see for example \cite{ellis,Buchert:2007ik,Kolb:2005da,Notari:2005xk,Coley:2005ei,Rasanen:2006kp,Kai:2006ws,Paranjape:2007wr,Wiltshire:2007jk,Kolb:2008bn,Bolejko:2008yj,Brown:2009tg,Larena:2008be,Clarkson:2009hr}.

This paper is organized as follows.
In Section \ref{model} we will introduce the specific $\Lambda$LTB model we will use, and in Section \ref{analysis} we will discuss how to confront the $\Lambda$LTB model against the observations.
In Section \ref{results} we will show our results.
We will see that the $\Lambda$CDM model is not the only possibility compatible with the observations, and that a matter-only void model is a viable alternative to the concordance model only if the BAO constraints are relaxed.
Moreover, we will show that the areas of the parameter space which give a good fit to the observations are always disconnected.
In particular, we will see that a local void does not affect the parameter extraction for $\Lambda$CDM models if its radius is smaller than 1-2~Gpc.
Finally, we will give our conclusions in Section \ref{outlook} and we will discuss a possible degeneracy between $\delta_{\Omega}$ and $\Omega_{K, \text{out}}$ in Appendix \ref{degen}.

Together with this paper we release the Mathematica package \mbox{\tt LLTB 1.0}, which is available at the address \mbox{\tt turbogl.org/LLTB.html}.

\section{The $\Lambda$LTB model}\label{model}

We will now quickly review the conventional LTB formalism.
The general picture is of a local LTB void exactly matched to the chosen Friedmann-Lema\^{i}tre-Robertson-Walker (FLRW) background.
For more details see, for example, Ref.~\cite{Enqvist:2006cg}.

\subsection{Basic formalism} \label{basif}

The line element of the spherically symmetric LTB model can be written as ($c=1$):
\begin{equation} \label{metric}
   \text{d}s^2=-\text{d}t^2+\frac{Y'^2(r,t)}{1-k(r)}\text{d}r^2+Y^2(r,t)(\text{d}\theta^2+\sin^2\theta\text{d}\phi^2) \,,
\end{equation}
where $Y(r,t)$ is the scale function, the prime denotes derivation with respect to the coordinate radius $r$ and the arbitrary function $k(r)$ represents the local curvature.
The FLRW solution is recovered by setting $Y(r,t)\rightarrow a(t) \, r^2$ and $k(r)\rightarrow k \, r^2$ throughout the equations.
Note that in the LTB space the transverse expansion rate $H_{T} \equiv \dot Y/Y$ will generally differ from the longitudinal expansion rate $H_{L} \equiv \dot Y'/Y'$.
For later use we also define the background expansion rate by $H_{\text{out}}(t) \equiv \dot a(t)/a(t)$.

The dynamics of the model is governed by the following equation \cite{Bondi:1947av}:
\begin{equation} \label{dynamics}
   \frac{\dot{Y}^2}{Y^2}=\frac{F(r)}{Y^3}+\frac{8\pi G}{3}\rho_\Lambda -\frac{k(r)}{Y^2} \,,
\end{equation}
where the dot denotes derivation with respect to the coordinate time $t$ and $\rho_\Lambda = \Lambda / 8\pi G$ is the energy density associated with the cosmological constant. The arbitrary function $F(r)$ (actually a constant of integration) represents the effective gravitating mass and is related to the local dust energy density $\rho_M(r,t)$ through the equation $F'= 8\pi G \, \rho_M \,Y'Y^2$.
It is useful to rewrite Eq.~(\ref{dynamics}) in the following, more familiar form
\begin{equation} \label{dynamics2}
   H_T^2(r,t)=H_0^2(r)\left[\Omega_M(r)\left(\frac{Y_0(r)}{Y(r,t)}\right)^3+\Omega_\Lambda(r)+\Omega_K(r)\left(\frac{Y_0(r)}{Y(r,t)}\right)^2\right] \,,
\end{equation}
where $H_0(r)\equiv H_T(r,t_0)$, $Y_0(r)\equiv Y(r,t_0)$ and the (present-day) density parameters are
\begin{eqnarray}
  \Omega_M(r) & \equiv & {F(r) \over  H_{0}^{2}(r) Y_{0}^{3}(r)}     \,, \label{omegamg}  \\
  \Omega_\Lambda(r) & \equiv & {8\pi G \over  3} {\rho_{\Lambda} \over  H^2_0(r)}    \,, \\
    \Omega_K(r) & \equiv & 1- \Omega_M(r) -   \Omega_\Lambda(r) = - {k(r) \over H^2_0(r) Y^2_0(r)} \,.
\end{eqnarray}
We point out that these density parameters have their usual meaning (ratios of energy densities over the critical density) only within the homogeneous patches (origin included) where $H_{L}=H_{T}$, whereas generally the local critical density is given by $\rho_{c}= (H_{T}^{2}+2 H_{T}H_{L})/8 \pi G$. 
For later use we define also the present-day expansion rate at the observer's position $H_{0, \text{in}} = H_0(r=0)$ and the background present-day expansion rate $H_{0, \text{out}} = H_0(r \gg r_{0})$, which clearly coincides with $H_{\text{out}}(t_{0})$.

Eq.~(\ref{dynamics2}) can be used to determine the age of the universe at a radial coordinate $r$:
\begin{equation} \label{age}
   t_0-t_{B}(r)=\frac{1}{H_0(r)}\int\limits^1_0 \frac{dx}{\sqrt{\Omega_M(r)x^{-1}+\Omega_K(r)+\Omega_{\Lambda}(r)x^2}}\;.
\end{equation}
We choose to constrain the models by requiring a simultaneous big bang, i.e., by setting $t_B'(r)=0$.
Furthermore, we set the moment of the initial singularity at $t=0$, so that $t_0$ is the actual age of the universe. Simultaneous big bang excludes decaying modes which would be strongly in contradiction with the inflationary paradigm \cite{Zibin:2008vj}.

\subsection{Specific model and its parameter space} \label{chosenmo}

LTB models feature three arbitrary functions. Within the present formalism they are taken as $\Omega_{M}(r)$ (or $F(r)$), $t_{B}(r)$ (or $k(r)$) and $Y_{0}(r)$.
One of these is but an expression of the gauge freedom, which we fix by setting $Y_0(r)=r$ and, consequently, $a(t_{0})=1$ where $a(t)$ is the scale factor of the background FLRW model.
Another free function, $t_{B}(r)$, was already set by the simultaneous big bang condition and so we are left with specifying the matter profile $\Omega_{M}(r)$.
We will adopt the parameterization of Ref.~\cite{GarciaBellido:2008nz} (constrained GBH) which reads: 
\begin{equation} \label{omegam}
   \Omega_M(r)=\Omega_{M, \text{out}}+(\Omega_{M, \text{in}}-\Omega_{M, \text{out}}) \, \frac{1-\tanh ( r-r_0 / 2\Delta r )}{1+\tanh (r_0/ 2\Delta r )} \,,
\end{equation}
where the parameters $r_0$ and $\Delta r$ characterize respectively size and steepness of the density profile, while $\Omega_{M, \text{in}}$ and $\Omega_{M, \text{out}}$ are the matter density parameters at the observer's position and in the FLRW background outside the void.

The precise form of the density profile should not be essential. The LTB void models (and void models in general) depend crucially on only two \emph{physical} parameters:
the void depth $\delta_{\Omega} \equiv (\Omega_{M, \text{in}}-\Omega_{M, \text{out}})/ \Omega_{M, \text{out}}$ which gives the jump $\Delta H$ in the expansion rate required to mimic the measured acceleration, and the void size $r_{0}$ which sets the redshift extension of the void.
Any strong dependence on the precise shape of the profile (in our case the steepness $\Delta r$) would in fact signal fine tuning.

We remind that our local LTB void is matched to the outside background FLRW model (where it is $Y= a\, r$) and so the local matter density profile exhibits a compensating overdense shell surrounding the central underdensity.
We also point out that, as one can see from Eq.~(\ref{omegamg}), $\Omega_{M, \text{in}}$ is calculated with respect to the local critical density, i.e., using the higher local expansion rate.
Therefore the void depth function $\delta_{\Omega}$ does not exactly correspond to the actual matter contrast $\delta_{M}$; for the parameter range we use in this work we typically find that $\delta_{M} \approx \delta_{\Omega}+0.1$.

Summarizing, the background FLRW model will be specified by the parameters $\Omega_{\Lambda, \text{out}}$, $\Omega_{K, \text{out}}$ and $t_{0}$, while the void is modelled by $\delta_{\Omega}$, $r_{0}$ and $\Delta r$.
Moreover, we will leave the spectral index $n_{s}$ free and so the overall parameter space will be seven dimensional.

\section{Cosmological data analysis} \label{analysis}

In this Section we will carefully explain how to compare the $\Lambda$LTB model predictions for an observer at its center with supernovae (SNe), Hubble constant, cosmic microwave background (CMB) and baryon acoustic oscillations (BAO) observations.
Before examining each of these datasets individually, we need to develop the formalism of the \emph{effective model} necessary to confront CMB and BAO data.

The LTB metric is matched to the background FLRW model at some radius $r$.
Because the LTB solution corresponds to a pressureless dust source we enforce, similarly to Ref.~\cite{Zibin:2008vk}, the matching to happen at a redshift at which radiation is still negligible.
The value $\bar z=100$ satisfies this requirement and allows the void to be large enough so as not to constrain artificially the parameter space.
The relevant physics leading to CMB and BAO features occurs at redshifts much greater than $\bar z$ and so it is possible to describe the relative light cone by means of an effective FLRW model.
One could be tempted to use the Einstein-de Sitter model for fitting because at $\bar z$ cosmological constant and curvature are locally negligible.
The curvature, however, does not enter only the Friedmann equation, but also the metric with the global effect of changing, for example, the angular diameter distance.
Because, differently from Ref.~\cite{Zibin:2008vk,Moss:2010jx}, we will consider background models with nonzero curvature, we will use, similarly to Ref.~\cite{Biswas:2010xm}, the FLRW background itself as the effective model.

The calculation proceeds as follows.
After finding the present-day expansion rate profile $H_{0}(r)$ which satisfies Eq.~(\ref{age})
(this must be done numerically because no analytic solution can be found for $H_{0}(r)$ when $\Lambda \neq 0$),
we solve the dynamics with Eq.~(\ref{dynamics2}).
We then solve the light cone using the following geodesic equations:
\begin{eqnarray}
   \frac{dt}{dz}&=&-\frac{Y'(r,t)}{(1+z)\dot{Y}'(r,t)} \,,  \label{lc1} \\
   \frac{dr}{dz}&=&\frac{\sqrt{1-k(r)}}{(1+z)\dot{Y}'(r,t)} \,,  \label{lc2}
\end{eqnarray}
with initial conditions $t(0)=t_{0}$, $r(0)=0$.
As it is clear from the metric (\ref{metric}), the angular diameter and luminosity distance are then simply given by:
\begin{eqnarray}
d_{A}(z) &=& Y(r(z),t(z)) \,, \\
d_{L}(z) &=& (1+z)^{2} \, d_{A}(z) \label{lumidi} \,.
\end{eqnarray}
The effective model and the LTB model have to give the same angular diameter distance $\bar d_{A}$ at the matching spacetime point $(\bar r, \bar t)=(r(\bar z), t (\bar z ))$.
To achieve this we place the effective observer at $r=0$ and we solve the light cone for the background FLRW model (ratio of Eqs.~(\ref{lc1}-\ref{lc2}) in the FLRW limit):
\begin{equation}
\frac{dt}{dr} = -  {a(t) \over  \sqrt{1-k \, r^{2}}} \,,
\end{equation}
with initial condition $t(\bar r)= \bar t$.
The angular diameter distances now coincide because outside the void the scale function $Y$ matches the FLRW scale factor : $a(\bar t) \, \bar r = Y(\bar r, \bar t) =\bar d_{A}$.

Finally, the effective metric is simply the background FLRW model at time $t(r=0)=t_{0, \text{eff}}$. To be explicit, it is specified by the following parameters:
\begin{eqnarray}
H_{0, \text{eff}}&=&H_{\text{out}}(t_{0, \text{eff}})   \,,  \label{h0e}  \\
T_{0, \text{eff}} &=&  {1+\bar z \over 1+ \bar z_{\text{eff}}} \, T_{0}   \,,  \\
\Omega_{\gamma, \text{eff}} &=&   2.469 \cdot 10^{-5} \, h_{\text{eff}}^{-2} \, \left( {T_{0, \text{eff}} \over T_{0}} \right )^{4}  \,, \label{omegage} \\
\Omega_{R, \text{eff}} &=& 1.692 \, \Omega_{\gamma, \text{eff}} \,, \\
\Omega_{\Lambda, \text{eff}} &=& {H^{2}_{0, \text{out}} \over H^{2}_{0, \text{eff}}} \, \Omega_{\Lambda, \text{out}} \,, \\
\Omega_{K, \text{eff}} &=& {H^{2}_{0, \text{out}} \over H^{2}_{0, \text{eff}}} \, {a^{2}(t_{0}) \over a^{2}(t_{0, \text{eff}})} \, \Omega_{K, \text{out}}   \,, \\
\Omega_{M, \text{eff}} &=& 1- \Omega_{R, \text{eff}}  - \Omega_{\Lambda, \text{eff}} - \Omega_{K, \text{eff}}  \,,  \\
\Omega_{B, \text{eff}} &=&  0.0226 \, h_{\text{eff}}^{-2} \, \left( {T_{0, \text{eff}} \over T_{0}} \right )^{3} \,. \label{omegabe}
\end{eqnarray}
The first two lines define the effective Hubble constant and the effective CMB temperature. The CMB temperature measured by the real observer is set to $T_{0}=2.725$ K and $\bar z_{\text{eff}} \equiv a(t_{0, \text{eff}})/ a(\bar t)-1$. The dimensionless Hubble constant $h_{\text{eff}}$ is, as usual, defined by $H_{0, \text{eff}}=100 \, h_{\text{eff}}$ km s$^{-1}$ Mpc$^{-1}$.
Eqs.~(\ref{omegage}-\ref{omegabe}) specify the effective density parameters; the modifications are due to the different time and CMB temperature experienced by the effective observer.
Eq.~(\ref{omegabe}) comes from requiring a fixed baryon-photon number density ratio \cite{Steigman:2006nf}: the void observer's value is fixed to the WMAP7 \cite{Jarosik:2010iu} result of $\Omega_{B}h^{2}=0.0226$, which is also compatible with Big Bang Nucleosynthesis constraints \cite{O'Meara:2006mj}.

We will typically find that the effective CMB temperature is close to the observed one $T_{0, \text{eff}} \simeq T_{0}$.
This is mainly due to the fact that we are considering compensated voids, whose metric matches the outside background metric.
The consequence is that for observables outside the LTB patch changes in redshift are small (see for example Ref.~\cite{Marra:2007pm}).
While this choice seems natural if one wants to recover a homogeneous universe on very large scales, it constrains the LTB model and its ability to fit the observables. This is similarly true for the simultaneous big bang condition we imposed with Eq.~(\ref{age}).
We will develop these ideas in our Conclusions.

\subsection{Supernovae observations}

We will use the recent Union2 SNe Compilation \cite{Amanullah:2010vv}, which consists 557 type Ia supernovae in the redshift range $z=0.015-1.4$.
The predicted magnitudes are related to the luminosity distance $d_{L}$ of Eq.~(\ref{lumidi}) by:
\begin{equation}
m(z)=5\log_{10}d_{L}(z)/10\,\textrm{pc} \,,
\end{equation}
and so the likelihood analysis is based on the $\chi^2$ function:
\begin{equation}
   \chi'^2_{SNe}=\sum_i \frac{[m_{i}-m(z_i)+\mu]^2}{\sigma_i^2} \,,
\end{equation}
where the index $i$ labels the Union2 entries.
The parameter $\mu$ is an unknown offset sum of the SNe absolute magnitudes, of $k$-corrections and other possible systematics.
As usual, we marginalize the likelihood $L'_{SNe}= \exp (-\chi'^2_{SNe}/2)$ over $\mu$, $L_{SNe}= \int d\mu \, L'_{SNe}$, leading to a new marginalized $\chi^2$ function:
\begin{equation}
\chi_{SNIa}^2 = S_2-\frac{S_1^2}{S_0} \,,
\end{equation}
where we dropped a cosmology-independent normalizing constant and the auxiliary $S_{n}$ is defined by:
\begin{equation}
S_n\equiv\sum_i\frac{(m_{i}-m(z_i))^n}{\sigma_i^2} \,.
\end{equation}

Note that since $\mu$ is degenerate with $\log_{10}H_{0, \text{in}}$ we are effectively marginalizing also over the observer's Hubble constant.
This is particularly important within the analysis of $\Lambda$LTB models where we can have models with very small but deep voids for which the value of the Hubble constant at the center is not as relevant in fitting the SNe as it is for large voids.

\subsection{Hubble constant}

The Hubble constant is obtained by measuring cosmological standard candles mostly within a distance of roughly $200$ Mpc \cite{Freedman:2000cf}.
We therefore compute the local Hubble constant $H_{\text{loc}}$ by averaging the expansion rate profile within a sphere of radius $r_{\text{loc}}=200$ Mpc:
\begin{equation} \label{hloco}
H_{\text{loc}}=   \int_{0}^{r_{\text{loc}}}  H_{0}(r) \, 4\pi r^{2} dr  \Big/  ({4 \pi / 3} \; r_{\text{loc}}^{3}) \,,
\end{equation}
where we used the fact that $Y_{0}(r)=r$ and we neglected the here irrelevant curvature.

This averaging is meaningful in general, but it is of particular importance for an exploration of $\Lambda$LTB models.
Indeed one can immagine a mixed scenario consisting of the concordance model plus a very small ($r_{0} \ll 200$ Mpc) but deep void.
The unaveraged expansion rate at the very origin $H_{0, \text{in}}$ would then differ significantly from the averaged $H_{\text{loc}}$ which better reflects the observable data being close to the background expansion rate $H_{0, \text{out}}$.

The present-day determination of $H_{\text{loc}}$ is converging but different groups still disagree on the precise value.
We will consider the following two results from Ref.~\cite{Sandage:2006cv} and Ref.~\cite{Riess:2009pu}:
\begin{eqnarray}
H_{\text{S06}}&=&62.3 \pm 5.2  \textrm{ km s}^{-1} \textrm{Mpc}^{-1} \phantom{ciaooo} \textrm{Sandage et al. 2006}  \,,  \label{S06} \\
H_{\text{R09}}&=&74.2 \pm 3.6  \textrm{ km s}^{-1} \textrm{Mpc}^{-1}  \phantom{ciaooo} \textrm{Riess et al. 2009} \,.  \label{R09} 
\end{eqnarray}
We will perform the likelihood analysis separately for the two results, defining the respective $\chi^2$ functions as:
\begin{eqnarray}
   \chi^2_{\text{S06}}&=&\frac{(H_{\text{S06}}-H_{\text{loc}})^2}{\sigma_{\text{S06}}^2}  \,, \\
   \chi^2_{\text{R09}}&=&\frac{(H_{\text{R09}}-H_{\text{loc}})^2}{\sigma_{\text{R09}}^2}  \,.
\end{eqnarray}
%

\subsection{Cosmic microwave background} \label{scmb}

We will use positions and amplitudes of peaks and troughs in the CMB spectrum to constrain the $\Lambda$LTB models.
The location of peaks and troughs can be parametrized as~\cite{Hu:2000ti}:
\begin{equation} \label{lmp}
   l_m=(m-\phi_m) \, l_A \,,
\end{equation}
where the peaks are labelled by integer values of $m$ and the troughs by half-integer values, the quantity $\phi_m$ is a phase-shift parameter determined by pre-recombination physics and $l_A$ is the acoustic scale which is given by:
\begin{equation}
   l_A=\pi \, \frac{d_{A}(z^{*}) (1+z^{*})}{r_s^*} \,,
\end{equation} 
where angular diameter distance, sound horizon $r_s^*$ and redshift $z^{*}$ at recombination are computed using the effective model of Eqs.~(\ref{h0e}-\ref{omegabe}).
In our analysis we will consider the position of the first, second, third peak and of the first trough.
We will compute the corresponding phases $\phi_{1}$, $\phi_{1.5}$, $\phi_{2}$ and $\phi_{3}$ using the accurate analytical fits of Ref.~\cite{Doran:2001yw}.
We will also consider the relative heights of second and third peak relative to the first one, $H_{2}$ and $H_{3}$, for which we can use the fits of Ref.~\cite{Hu:2000ti}.

We constrain the $\Lambda$LTB model by means of the following $\chi^2$ function built from the six quantities explained above:
\begin{eqnarray} \label{chi2cmb}
   \chi^2_{\text{WMAP}}      &=&  
    \frac{(l_1-l_{1, \text{W7}})^2}{\sigma_{l_1}^2}  +  \frac{(l_{1.5}-l_{1.5, \text{W7}})^2}{\sigma_{l_{1.5}}^2} +\frac{(l_2-l_{2, \text{W7}})^2}{\sigma_{l_2}^2}  \nonumber \\
   &+& \frac{(l_3-l_{3, \text{W7}})^2}{\sigma_{l_3}^2}   +\frac{(H_2-H_{2, \text{W7}})^2}{\sigma_{H_2}^2}  +  \frac{(H_3-H_{3, \text{W7}})^2}{\sigma_{H_3}^2}   \,,
\end{eqnarray}
where the values marked with W7 correspond to the best-fit WMAP7 spectrum~\cite{Jarosik:2010iu}.
The error has to take into account both the experimental error and the error due to the fit.
We thought it reasonable to use a $\sigma$ of 1\% for the position of the first peak and of 3\% for the other quantities.
We checked the accuracy of the fits by comparing their predictions with the ones of CAMB~\cite{Lewis:1999bs} within our parameter space.

We would like to stress that the quantity $\chi^2_{\text{WMAP}}$ is a function of the spectral index $n_{s}$.
In particular, because it is computed through analytical fits, it is an analytical function over which we can easily integrate.
We will see in Section \ref{results} the importance of marginalizing over the spectral index.

\subsection{Baryon acoustic oscillations}

A baryon acoustic peak is detected from SDSS and 2dFGRS galaxy catalogues at redshifts $0.2$ and $0.35$ \cite{Percival:2007yw}.
The observed quantity is $\Delta\theta^2\Delta z$, where $\Delta\theta$ is the angle that the acoustic scale subtends on the sky in the transverse direction at the observed redshift, and $\Delta z$ is the redhisft interval corresponding to the acoustic scale in the radial direction.
In order to test void models against the BAO results, we have to understand how to compute the quantity $\Delta\theta^2\Delta z$ in a void model. This procedure is explained in detail in Ref.~\cite{Biswas:2010xm}.

First we need the comoving acoustic scale at the drag epoch $r_s^{\text{drag}}$, which is defined as the time at which the baryons are released from the Compton drag of the photons.
Generally, in an inhomogeneous universe the drag epoch corresponds to different times in different locations: $t_{d}=t_d(r)$. However, because we constrained the LTB model to have a simultaneous big bang (see Section \ref{basif}), the universe is almost homogeneous at the drag epoch and so the $r$ dependence in $t_d(r)$ is very weak.
$r_s^{\text{drag}}$ can then be computed using the effective metric and the fitting formulas of Ref.~\cite{Eisenstein:1997ik}.
The latter, however, are tuned for models with a CMB temperature of $T_{0}=2.725$ K and so we have to correct the relevant quantities with the appropriate powers of $T_{0} / T_{0, \text{eff}}$.

Because in the LTB metric transverse and longitudinal expansion rates differ ($H_{T} \neq H_{L}$), the acoustic scale is stretched differently in the transverse and radial directions. The correct scaling is given by the following formulas:
\begin{eqnarray}
   l_T(z) & = & \frac{Y(r(z),t(z))}{Y(r(z),t_d(r))} \, l_s^{\text{drag}}(z)  \,, \\
   l_L(z) & = & \frac{Y'(r(z),t(z))}{Y'(r(z),t_d(r))} \, l_s^{\text{drag}}(z)  \,,  
\end{eqnarray}
where $l_s^{\text{drag}}(z)$ is the proper acoustic scale at the drag epoch for the shell $r(z)$.
As for $t_d(r)$, the $z$ dependence in $l_s^{\text{drag}}(z)$ is very weak because our early universe was close to homogeneity and we can simply write:
\begin{equation} \label{lsd}
   l_s^{\text{drag}}=\frac{r_s^{\text{drag}}}{1+z_d} \,,
\end{equation}
where $z_{d}$ is calculated using the effective metric.

It is also interesting to see how the calculation proceeds in the general case where one cannot assume a weak $r$ dependence in $t_d(r)$.
The final result for $\Delta\theta^2\Delta z$ is indeed formally the same.
We start by noting (this can be seen from Eq.~(\ref{dynamics2})), that radial shells evolve as independent FLRW models, and we assume that $\Delta z$ is small enough so that the spherical shells at $r(z)$ and $r( z+\Delta z)$ can be described by the same FLRW model at the time $t( z)$, where $z$ is the redshift corresponding to the BAO observations.
Next we evolve this FLRW model to find an observer that would see the event at time $t(z)$ with the same redshift $z$ and we use this observer's cosmological parameters to calculate $r_s^{\text{drag}}$ and $z_{d}$ with the formulas in Ref.~\cite{Eisenstein:1997ik}, and then use Eq.~(\ref{lsd}) to find $l_s^{\text{drag}}$.

The quantities $\Delta \theta$ and $\Delta z$ can then be written as:
\begin{eqnarray}
   \Delta\theta(z) & = & \frac{l_T(z)}{d_A(z)}=\frac{l_T(z)}{Y(r(z),t(z))} \,, \\
   \Delta z(z)     & = & l_L(z) \, (1+z) \, H_{L}(r(z),t(z)) \,,
\end{eqnarray}
where in the last line we have used Eq.~(\ref{lc2}) in the limit $k(r),\, \Delta z\ll 1$.
Finally, we need the general relation:
\begin{equation} \label{zzz}
   1+ z_d=(1+z)\left(\frac{Y'(r(z),t(z)) \, Y^2(r(z),t(z))}{Y'(r(z),t_d(r)) \, Y^2(r(z),t_d(r))}\right)^{1/3} \,,
\end{equation}
which shows that the volume element at $r(z)$ has diluted by a factor $(1+z_d)/(1+z)$ between $t_d(r)$ and $t(z)$, i.e., that the density at drag epoch is correctly scaled.
Lastly, we can construct the quantity $(z^{-1}\Delta\theta^2\Delta z)^{1/3}\equiv r_s^{\text{drag}}/d_V(z)$ by identifying
\begin{equation}
   d_V(z)=\left( {z \, (1+z)^2 \, Y^2(r(z),t(z))  \over H_L(r(z),t(z))} \right)^{1/3}   \,,
\end{equation}
where we have used Eq.~(\ref{zzz}).
Note that the redshift $z_{d}$ has canceled in the final expression; this would not be the case if one were to consider $\Delta z$ and $\Delta\theta$ separately.

We will compare the void models with the results of Ref.~\cite{Percival:2007yw}: $r_s^{\text{drag}}/d_V(0.2)=0.1980\pm0.0058$ and $r_s^{\text{drag}}/d_V(0.35)=0.1094\pm0.0033$, by computing the $\chi^2$ from the multivariate Gaussian likelihood. That is with
\begin{equation}
   \chi^2_{BAO}=X^TV^{-1}X \,,
\end{equation}
where
$$
X=
\begin{pmatrix}
   \frac{r_s(z_d)}{d_V(0.2)}-0.1980 \\
   \frac{r_s(z_d)}{d_V(0.35)}-0.1094 
\end{pmatrix}
~~\text{and}~~
V^{-1}=
\begin{pmatrix}
   35059  & -24031 \\
   -24031 & 108300
\end{pmatrix} \,.
$$

\section{Results} \label{results}

As summarized at the end of Section \ref{model}, the parameter space of our $\Lambda$LTB model is seven dimensional and consists of the parameters
$\Omega_{\Lambda, \text{out}}$, $\Omega_{K, \text{out}}$, $t_{0}$ and $n_{s}$ for the background FLRW model and $\delta_{\Omega}$, $r_{0}$ and $\Delta r$ for the void.
In this analysis we want to focus on (possibly) matter-only voids and so we set  $\delta_{\Omega}=-0.9$ (which corresponds to a matter contrast of $\delta_{M}\approx -0.81$) and $\Delta r = 0.35 \, r_{0}$, because this choice always guarantees a good fit to the SNe.
We will explore in Appendix \ref{degen} different values of $\delta_{\Omega}$. In particular, we will show that there is a degeneracy between $\delta_{\Omega}$ and $\Omega_{K, \text{out}}$ as far as a good fit to the SNe is concerned.

We are left with five parameters.
We choose $\Omega_{\Lambda, \text{out}}$ and $r_{0}$ as the parameters with respect to which to plot the likelihood surfaces. These parameters indeed best characterize the two opposite alternatives of the $\Lambda$CDM and matter-only void models, which delimit the parameter space.
We stress that this analysis will compare the two alternatives in an unbiased manner, thus overcoming some of the ambiguities in comparing the two models.
We remind that the LTB model is specified not by free parameters, but by free functions and therefore the number of degrees of freedom to be adopted in computing the reduced~$\chi^{2}$, which is usually used in model ranking, is necessarily a somewhat subjective choice.

Only three parameters are now left to specify: $n_{s}$, $t_{0}$ and $\Omega_{K, \text{out}}$.
We will plot in the indicated figures confidence level contours for:
\begin{eqnarray}
&&\textrm{Fig.~\ref{fig1}} \qquad L(r_0, \Omega_{\Lambda, \text{out}}, \bar \Omega_{K, \text{out}},\bar t_0, \bar n_{s})  \,, \label{li1} \\
&&\textrm{Fig.~\ref{fig2}} \qquad L(r_0, \Omega_{\Lambda, \text{out}}, \bar \Omega_{K, \text{out}},\bar t_0)  =
\int d n_{s}   \,  L(r_0, \Omega_{\Lambda, \text{out}}, \bar \Omega_{K, \text{out}}, \bar t_0, n_{s})  \,,  \label{li2} \\
&&\textrm{Fig.~\ref{fig3}} \qquad L(r_0, \Omega_{\Lambda, \text{out}}, \bar \Omega_{K, \text{out}})  =
\int d n_{s}   \,   d t_{0} \,  L(r_0, \Omega_{\Lambda, \text{out}}, \bar \Omega_{K, \text{out}}, t_0, n_{s})  \,,  \label{li3} \\
&&\textrm{Fig.~\ref{fig4}} \qquad L(r_0, \Omega_{\Lambda, \text{out}})   =
\int dn_{s}  \,   d t_{0} \, d \Omega_{K, \text{out}} \, L(r_0, \Omega_{\Lambda, \text{out}}, \Omega_{K, \text{out}}, t_0, n_{s}) \,,  \label{li4} \\
&&\textrm{Fig.~\ref{fig5}} \qquad L(r_0, \Omega_{\Lambda, \text{out}}, \bar n_{s})   =
\int d t_{0} \, d \Omega_{K, \text{out}}        \, L(r_0, \Omega_{\Lambda, \text{out}}, \Omega_{K, \text{out}}, t_0, \bar n_{s}) \,,  \label{li5} 
\end{eqnarray}
where $\bar n_{s}$, $\bar t_{0}$ and $\bar \Omega_{K, \text{out}}$ are some fixed values.
Throughout the analysis we will adopt the prior $r_{0} \le 3.5$ Gpc: larger values of the void radius would result in a void whose light cone extends till times at which radiation cannot be neglected (see the discussion about the effective metric in Section \ref{analysis}).
We will now show in Section \ref{wmapun} an illustrative example.
We will then present our main results in Sections \ref{nsuni}-\ref{nons}.

\subsection{Illustrative example} \label{wmapun}

\begin{figure}
\begin{center}
\includegraphics[width= .49\textwidth]{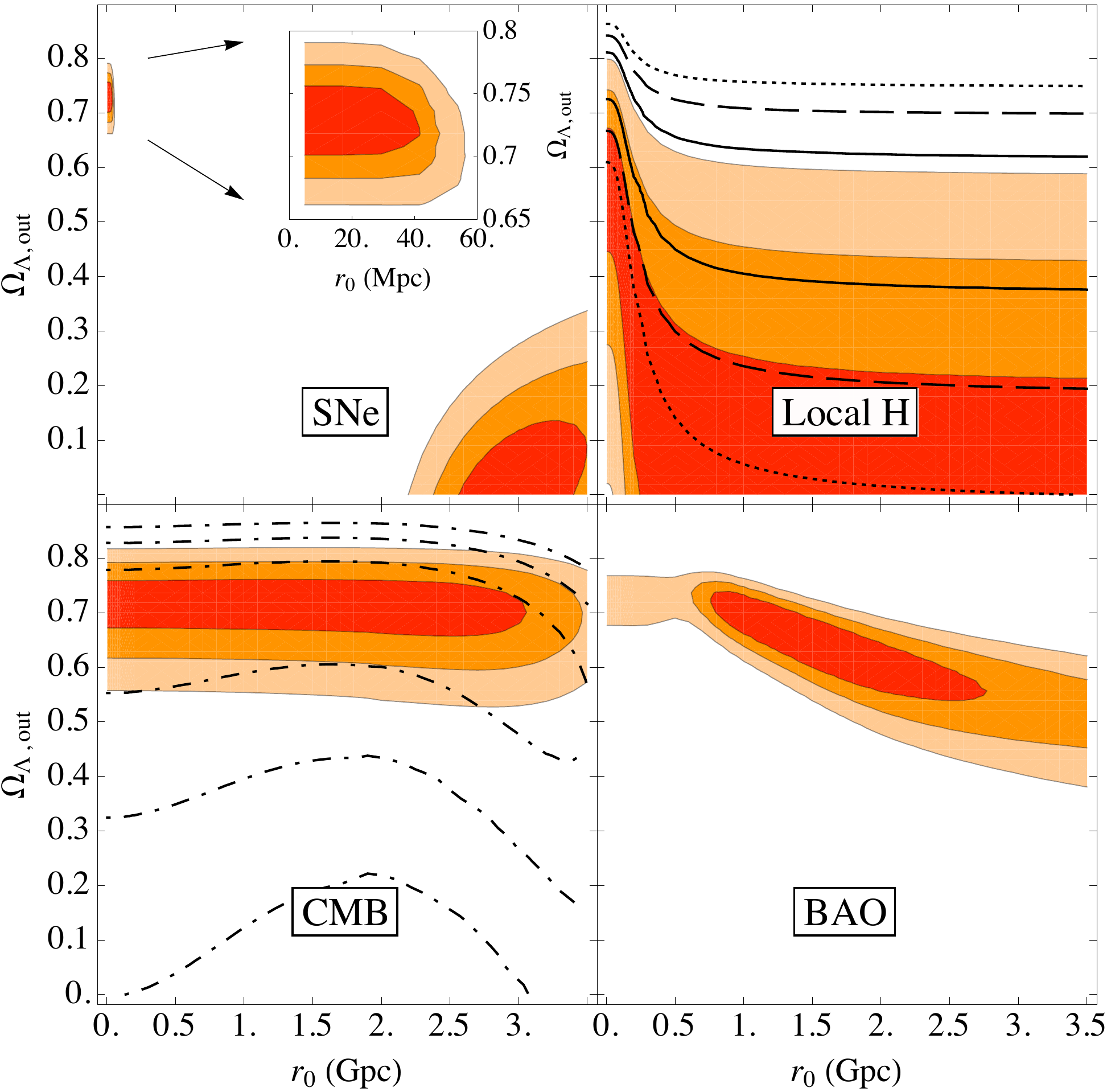}
\includegraphics[width= .49\textwidth]{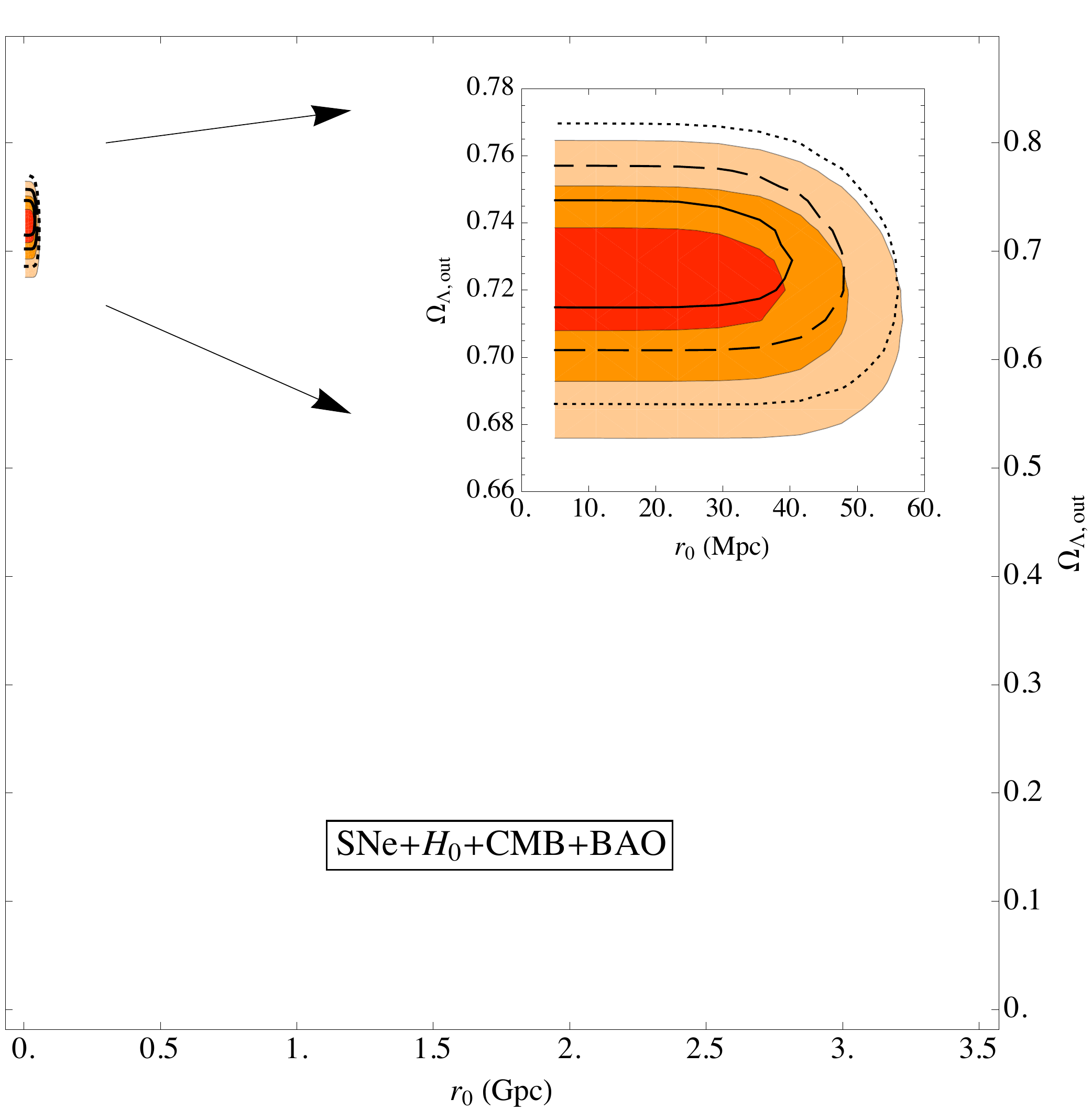}
\caption{
1, 2 and 3$\sigma$ confidence level contours on $r_{0}$ and $\Omega_{\Lambda, \text{out}}$ for the likelihood of Eq.~(\ref{li1}) for the $\Lambda$LTB model described in the text with $n_{s}=0.96$, $t_{0}=13.7$ Gyr and $\Omega_{K, \text{out}}=0$.
The four smaller panels on the left show the contours for the independent likelihoods per observable, while the larger panel on the right shows the contours for the combined likelihood.
In the panel representing measurements of the local Hubble constant, the results relative to $H_{\text{S06}}$ of Eq.~(\ref{S06}) are shown as filled contours, while the ones relative to $H_{\text{R09}}$ of Eq.~(\ref{R09}) are shown as lines. The same labelling holds for the panel relative to the combined observables.
In the panel relative to CMB contraints we also show confidence levels for the likelihood of Eq.~(\ref{li2}) marginalized over $n_{s}$ (dot-dashed contours).
See Section \ref{wmapun} for a discussion.
}
\label{fig1}
\end{center}
\end{figure}

We start by choosing $n_{s}=0.96$, $t_{0}=13.7$ Gyr and $\Omega_{K, \text{out}}=0$, which are the WMAP7 values corresponding to the standard $\Lambda$CDM cosmology \cite{Jarosik:2010iu}.
The results of the likelihood analysis are shown in Fig.~\ref{fig1}.
The four small panels on the left show the 1, 2 and 3$\sigma$ confidence level contours for the observables examined in the previous Section. We will now discuss them in detail.

The \emph{SNe} panel shows the known result that a pure-matter void model is as successful as the concordance $\Lambda$CDM model in fitting SNe observations.
Moreover, the present analysis of $\Lambda$LTB models shows that a mixed scenario of a $\Lambda$CDM model together with a large deep void is not favoured.
For smaller contrasts (as explained above we have fixed $\delta_{\Omega}=-0.9$) the contours corresponding to the void model move towards the ones of the concordance model.
However, the area of the parameter space with small $r_{0}$ and $\Omega_{\Lambda, \text{out}}$ (roughly delimited by the line joining concordance model and matter-only void) is excluded.
For example, a shallow void with $\delta_{\Omega}=-0.4$ and $\Omega_{\Lambda, \text{out}}=0.6$ still requires a radius of at least 1 Gpc.

The \emph{``Local H''} panel shows the constraints coming from local measurements of the Hubble constant.
We remind the reader that we have fixed the age of the universe $t_{0}$ and so the background expansion rate $H_{0, \text{out}}$ increases along the $\Omega_{\Lambda, \text{out}}$ axis.
To understand the shape of the contours it is useful to look back at Eq.~(\ref{hloco}).
If the void is large, then $H_{\text{loc}}$ will be higher than the background value $H_{0, \text{out}}$ because of the $\Delta H$ jump caused by $\delta_{\Omega}$.
If, however, the void radius is small, then the averaging of Eq.~(\ref{hloco}) gives back the lower $H_{0, \text{out}}$ value.
The latter trend, together with the fact that $H_{0, \text{out}}$ increases with $\Omega_{\Lambda, \text{out}}$, shows that the $H_{\text{loc}}$ relative to a point in the parameter space can also be obtained for smaller voids and higher $\Omega_{\Lambda, \text{out}}$.
For the parameters chosen, a matter-only model is consistent with the $H_{\text{S06}}$ of Eq.~(\ref{S06}), while it is not with the $H_{\text{R09}}$ of Eq.~(\ref{R09}).

The \emph{CMB} panel shows the constraints coming from the CMB spectrum.
We show both the confidence levels relative to $n_{s}=0.96$ (filled contours and Eq.~(\ref{li1})) and the ones for the likelihood marginalized over $n_{s}\in [0.8,1.2]$ (dot-dashed contours and Eq.~(\ref{li2})).
The fits we are using are valid in the latter range, which includes the values relevant for the likelihood analysis. 
For the parameters chosen the concordance $\Lambda$CDM model, possibly with a large local void, is favoured in both cases.
In the non marginalized case the contours depend weakly on $r_{0}$ because the effect of the void on the angular diameter distance is generally small for observables outside the LTB patch.
The dependence on $r_{0}$ becomes stronger for very large voids which extend deep into the past light cone.

The \emph{BAO} panel shows the constraints coming from the measurements of the acoustic oscillations in the matter power spectrum.
For this observable, the concordance model does not give a good fit, even though it is within the 3$\sigma$ confidence level contour.
It is interesting to see that a 1$\sigma$ confidence level fit is achieved by a $\Lambda$CDM model with $\Omega_{\Lambda, \text{out}}= 0.55 - 0.75$ together with a void of radius $r_{0} =0.8 - 2.7$ Gpc.
For the chosen $t_{0}$ and $\Omega_{K, \text{out}}$, a pure-matter void model is excluded.
We point out that by fine tuning the void profile beyond the Ansatz of Eq.~(\ref{omegam}) it is possible to fit the BAO data without cosmological constant~\cite{Biswas:2010xm}.

Finally, the large panel on the right in Fig.~\ref{fig1} shows the combined likelihood (for $n_{s}=0.96$) with the result that the standard concordance model is favoured against a void model of large radius of any type for these parameters. As shown in the inset, a small local void of few tens Mpc is not, however, excluded.
Note in particular that the confidence level contours are perpendicular to the $\Omega_{\Lambda, \text{out}}$ axis and so a small local void does not bias the parameter extraction within the framework of $\Lambda$CDM models.
See Ref.~\cite{Sinclair:2010sb} for an alternative analysis.

It is important to point out that the previous results depend crucially on the values chosen for $n_{s}$, $t_{0}$ and $\Omega_{K, \text{out}}$.
We decide, therefore, to marginalize the total likelihood over the latter parameters in order to analyze the data with as few as possible priors.
We show these results in the next Sections, while the findings of this Section should be considered as illustrative ones.
We stress that in this way we do not need either the reduced $\chi^{2}$ or the minimum $\chi^{2}$ to compare the $\Lambda$CDM with void models.

\subsection{Marginalizing over $n_{s}$} \label{nsuni}

\begin{figure}
\begin{center}
\includegraphics[width= .49\textwidth]{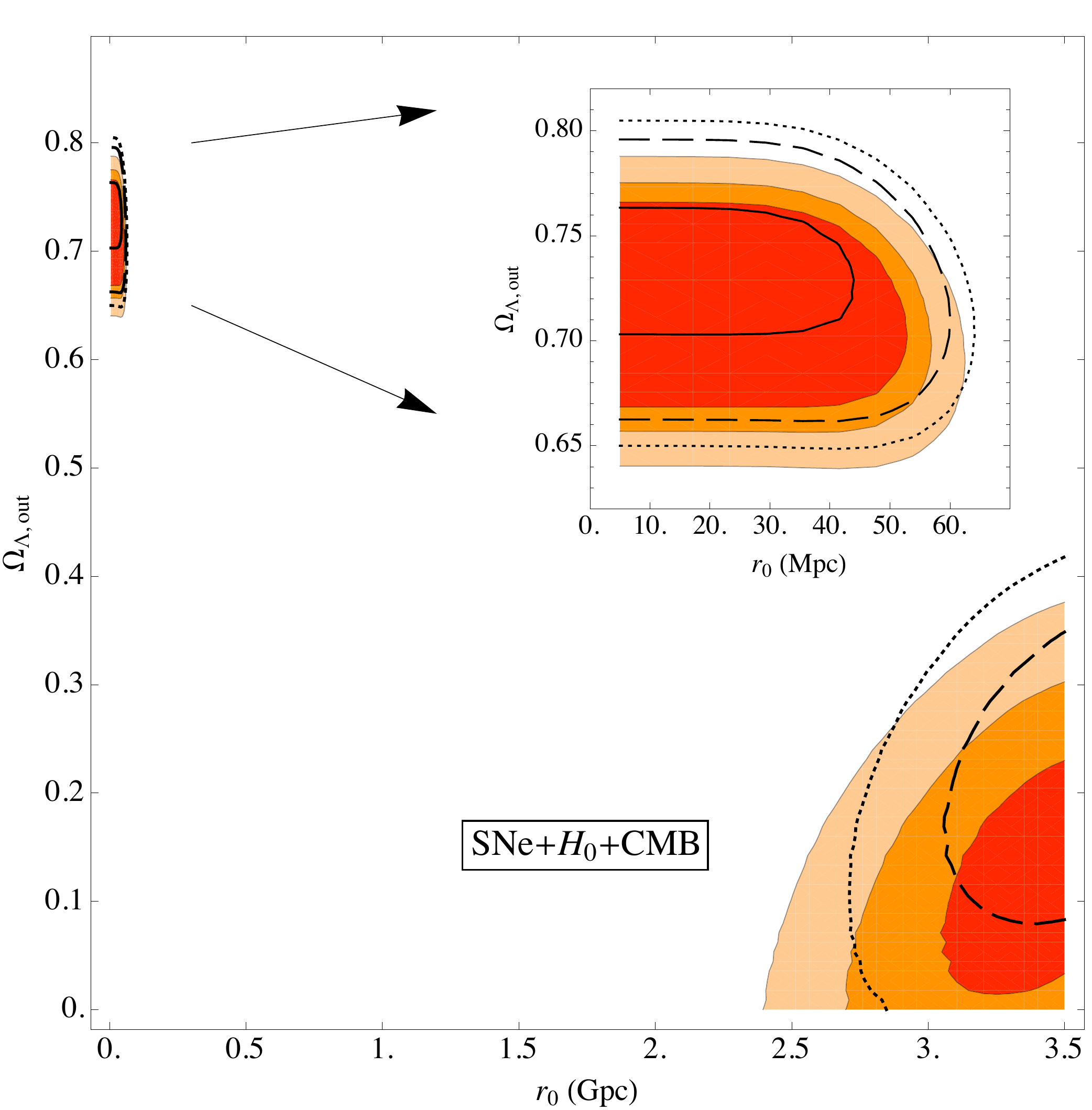}
\includegraphics[width= .49\textwidth]{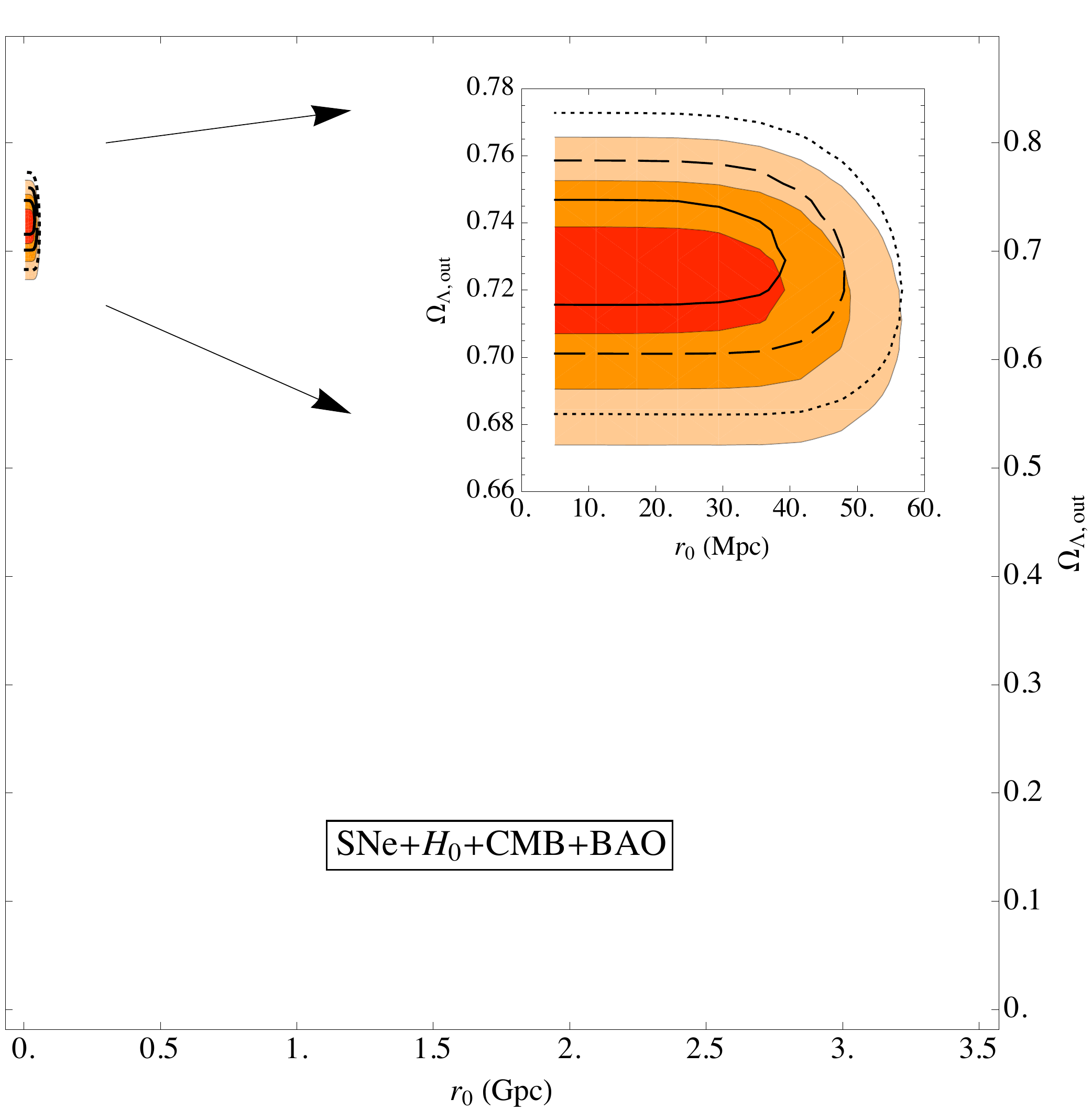}
\caption{
1, 2 and 3$\sigma$ confidence level contours on $r_{0}$ and $\Omega_{\Lambda, \text{out}}$ for the likelihood of Eq.~(\ref{li2}) for the $\Lambda$LTB model described in the text marginalized over $n_{s}$ and with $t_{0}=13.7$ Gyr and $\Omega_{K, \text{out}}=0$. Labelling as in Fig.~\ref{fig1}.
The right panel shows the combined likelihood for all observables, while the BAO constraints are excluded in the left panel.
See Section \ref{nsuni} for a discussion.
}
\label{fig2}
\end{center}
\end{figure}

In this Section we will consider models with $t_{0}=13.7$ Gyr and $\Omega_{K, \text{out}}=0$ but of any spectral index, that is, we marginalize the total likelihood over $n_{s}$ as shown in Eq.~(\ref{li2}).
Numerically we consider a range $n_{s}\in [0.8,1.2]$, which includes the values relevant for the likelihood analysis, whose results we plot in Fig.~\ref{fig2}.
The right panel corresponds to a combined likelihood for all observables with the result that the concordance model is favoured at 3$\sigma$ confidence level against a void model of large radius of any type.

In deriving the BAO contraints we assumed that the BAO scale is comoving. Whether this really is a good approximation is still unclear~\cite{Clarkson:2009sc}. However, it is difficult to improve the treatment of the BAO feature because the perturbation theory in the LTB background is not yet fully understood.
Therefore, in the left panel of Fig.~\ref{fig2} we repeated the analysis excluding the BAO contraints.
In the case of the likelihood with $H_{\text{S06}}$ of Eq.~(\ref{S06}), the 1$\sigma$ confidence level now contains, besides the concordance model, an (almost) pure-matter void model.
In the case of the likelihood with  $H_{\text{R09}}$ of Eq.~(\ref{R09}) only the 3$\sigma$ confidence level contains the pure-matter void model.

It is clear from our results that the areas of the parameter space which give a good fit to observations are always disconnected.
Moreover, as pointed out in the previous Section, the confidence level contours relative to the concordance model are perpendicular to the $\Omega_{\Lambda, \text{out}}$ axis.
These two facts combined together show that a local void does not affect the parameter extraction for the $\Lambda$CDM model if its radius is smaller than 1-2~Gpc. The same conclusions hold also for the results of the next Sections.

\subsection{Marginalizing over $n_{s}$ and $t_{0}$} \label{funi}

\begin{figure}
\begin{center}
\includegraphics[width= .49\textwidth]{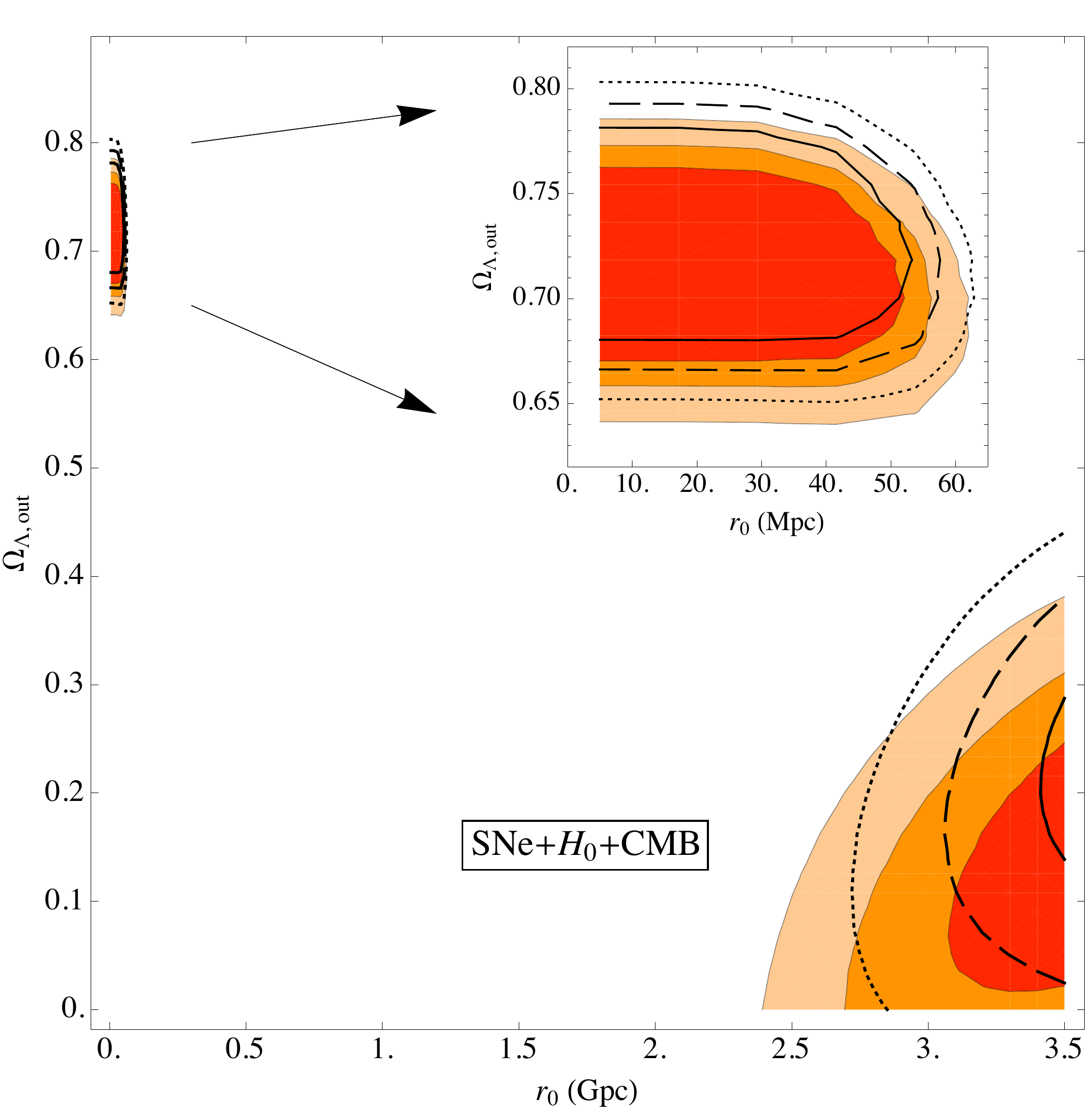}
\includegraphics[width= .49\textwidth]{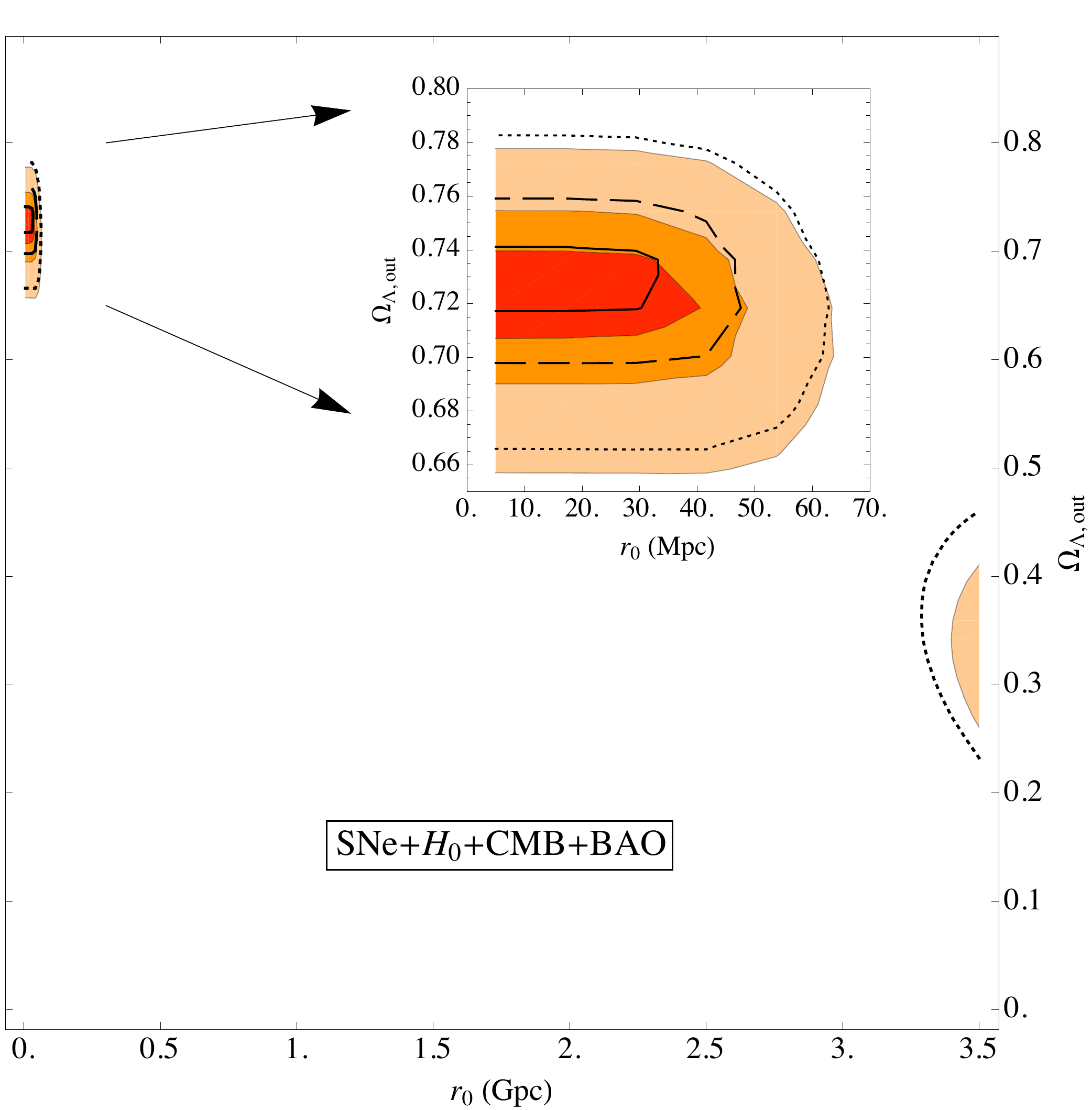}
\caption{
1, 2 and 3$\sigma$ confidence level contours on $r_{0}$ and $\Omega_{\Lambda, \text{out}}$ for the likelihood of Eq.~(\ref{li3}) for the $\Lambda$LTB model described in the text marginalized over $n_{s}$ and $t_{0}$ and with $\Omega_{K, \text{out}}=0$. Labelling and definitions as in Fig.~\ref{fig2}.
See Section \ref{funi} for a discussion.
}
\label{fig3}
\end{center}
\end{figure}

In this Section we will consider models with $\Omega_{K, \text{out}}=0$ but of any age and spectral index, that is, we marginalize the total likelihood over $t_{0}$ and $n_{s}$ as shown in Eq.~(\ref{li3}).
Observations of stellar globular clusters are reported to result in a lower bound of $t_{0}=11.2$ Gyr \cite{Krauss:2003em}. This bound, however, uses a $\Lambda$CDM template in order to relate the age of the oldest globular cluster in the Milky Way (10.4 Gyr) to the age of the universe.
Relaxing the assumption of the background model we obtain a slighter lower bound of $t_{0} \simeq 11$ Gyr.
We then take $t_{0}=16.7$ Gyr as upper bound, which we have found to be numerically equivalent to an unbounded upper limit as the total likelihood is already essentially zero.

The results are shown in Fig.~\ref{fig3} where, similarly to the previous Section, we include (right panel) and exclude (left panel) BAO constraints in the analysis.
With the BAO constraints the concordance model is favoured at 1 and 2$\sigma$ confidence level against a void model of large radius of any type.
At 3$\sigma$ confidence level, however, a mixed scenario with a void of radius $r_{0}=3.25-3.5$ Gpc and $\Omega_{\Lambda, \text{out}}=0.25-0.45$ is not excluded.
Without the BAO constraints the results are similar to the ones of the previous Section if $H_{\text{S06}}$ is used.
The contours are instead different in the case of the likelihood with $H_{\text{R09}}$, which at 1$\sigma$ confidence level now has a mixed scenario with a void of large radius and $\Omega_{\Lambda, \text{out}}=0.15-0.30$.
The concordance model is always within the 1$\sigma$ confidence level contour.

\subsection{Marginalizing over $n_{s}$, $t_{0}$ and $\Omega_{K, \text{out}}$} \label{unicu}

\begin{figure}
\begin{center}
\includegraphics[width= .49\textwidth]{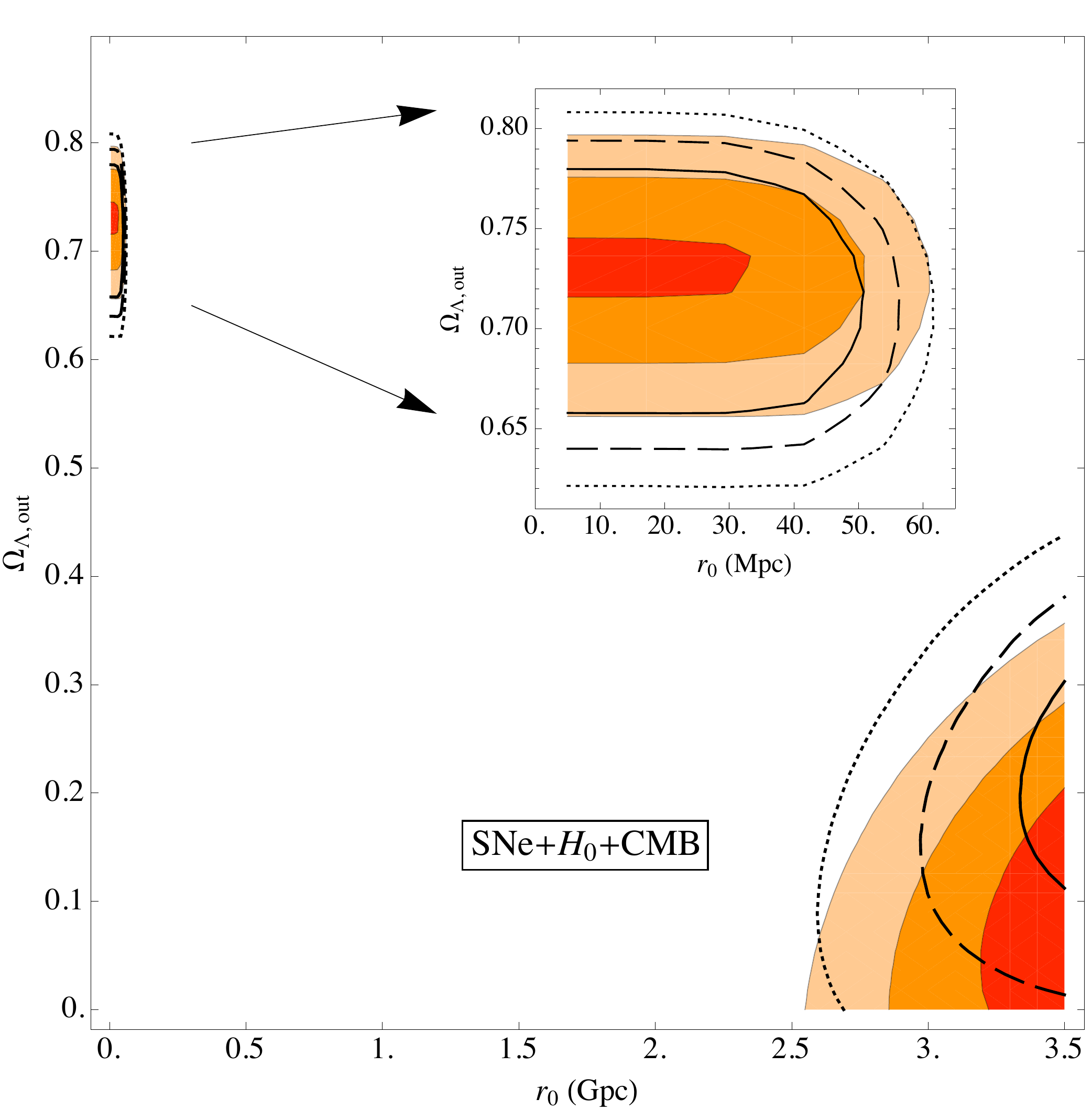}
\includegraphics[width= .49\textwidth]{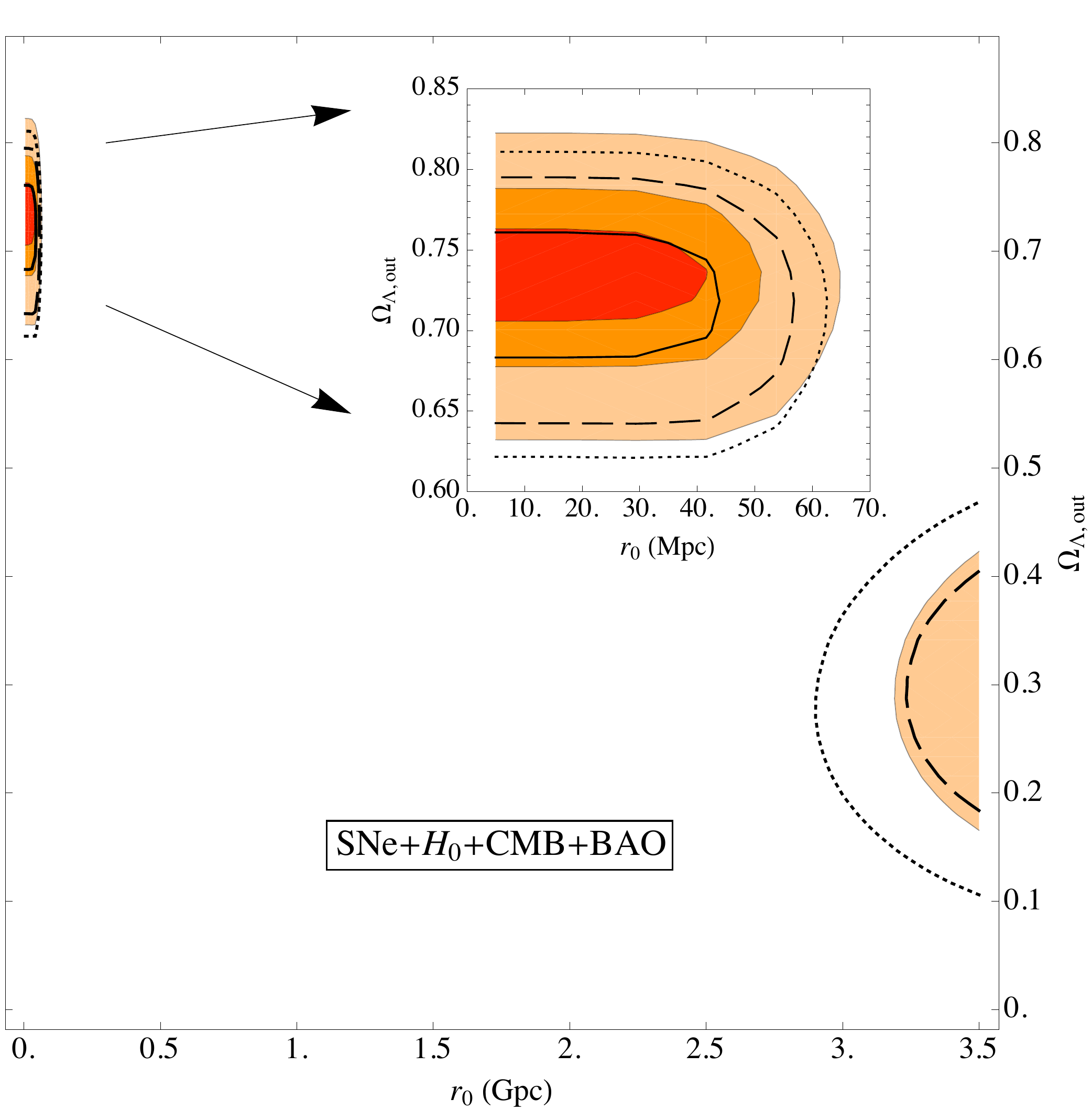}
\caption{
1, 2 and 3$\sigma$ confidence level contours on $r_{0}$ and $\Omega_{\Lambda, \text{out}}$ for the likelihood of Eq.~(\ref{li4}) for the $\Lambda$LTB model described in the text marginalized over $n_{s}$, $t_{0}$ and $\Omega_{K, \text{out}}$. Labelling and definitions as in Fig.~\ref{fig2}.
See Section \ref{unicu} for a discussion.
}
\label{fig4}
\end{center}
\end{figure}

In this Section we will consider models of  any spectral index, age and background curvature, that is, we marginalize the total likelihood over $n_{s}$, $t_{0}$ and $\Omega_{K, \text{out}}$ as shown in Eq.~(\ref{li4}).
As in the previous Sections, $n_{s}\in [0.8,1.2]$ and $t_{0} \in [11,\, 16.7]$ Gyr.
The background curvature range we consider is $\Omega_{K, \text{out}} \in [-0.25,\, 0.16]$ which includes the values relevant for the likelihood analysis.
The results are shown in Fig.~\ref{fig4} where again we include (right panel) and exclude (left panel) BAO constraints in the analysis.

Without the BAO contraints, the 1$\sigma$ confidence level fully includes the matter-only void model if $H_{\text{S06}}$ is used, while the concordance model is always included.
If $H_{\text{R09}}$ is used, the pure-matter void model is included only by the 3$\sigma$ confidence level.
With BAO constraints included the 1$\sigma$ confidence level always contains the concordance model but not the matter-only void model.
However, the 3$\sigma$ (filled contours relative to $H_{\text{S06}}$) or 2-3$\sigma$ (empty contours relative to $H_{\text{R09}}$) confidence levels include a mixed scenario of a large void with $\Omega_{\Lambda, \text{out}} \approx 0.1-0.5$.

We would like to point out that without the BAO constraints the matter-only void model can give as good a minimum $\chi^2$ as the concordance model if we use $H_{\text{S06}}$ for the local Hubble parameter.
For example, a pure-matter void model with $t_{0}=15.45$ Gyr and $\Omega_{K, \text{out}}=-0.1$ gives a minimum $\chi^2$ of 559.8, while the concordance model a $\chi^2$ of 559.4, where in both cases we marginalized over $n_{s}$.

We would like to comment that, within our modelling, a curved background can have two different interpretations.
$\Omega_{K, \text{out}}$ can indeed be relative to the global universe or it could just describe a patch bigger than the LTB void.
In the latter case one could imagine the presence of many patches of different $\Omega_{K, \text{out}}$ so that on average the universe is still flat.

\subsection{Marginalizing over $t_{0}$ and $\Omega_{K, \text{out}}$} \label{nons}

\begin{figure}
\begin{center}
\includegraphics[width= .49\textwidth]{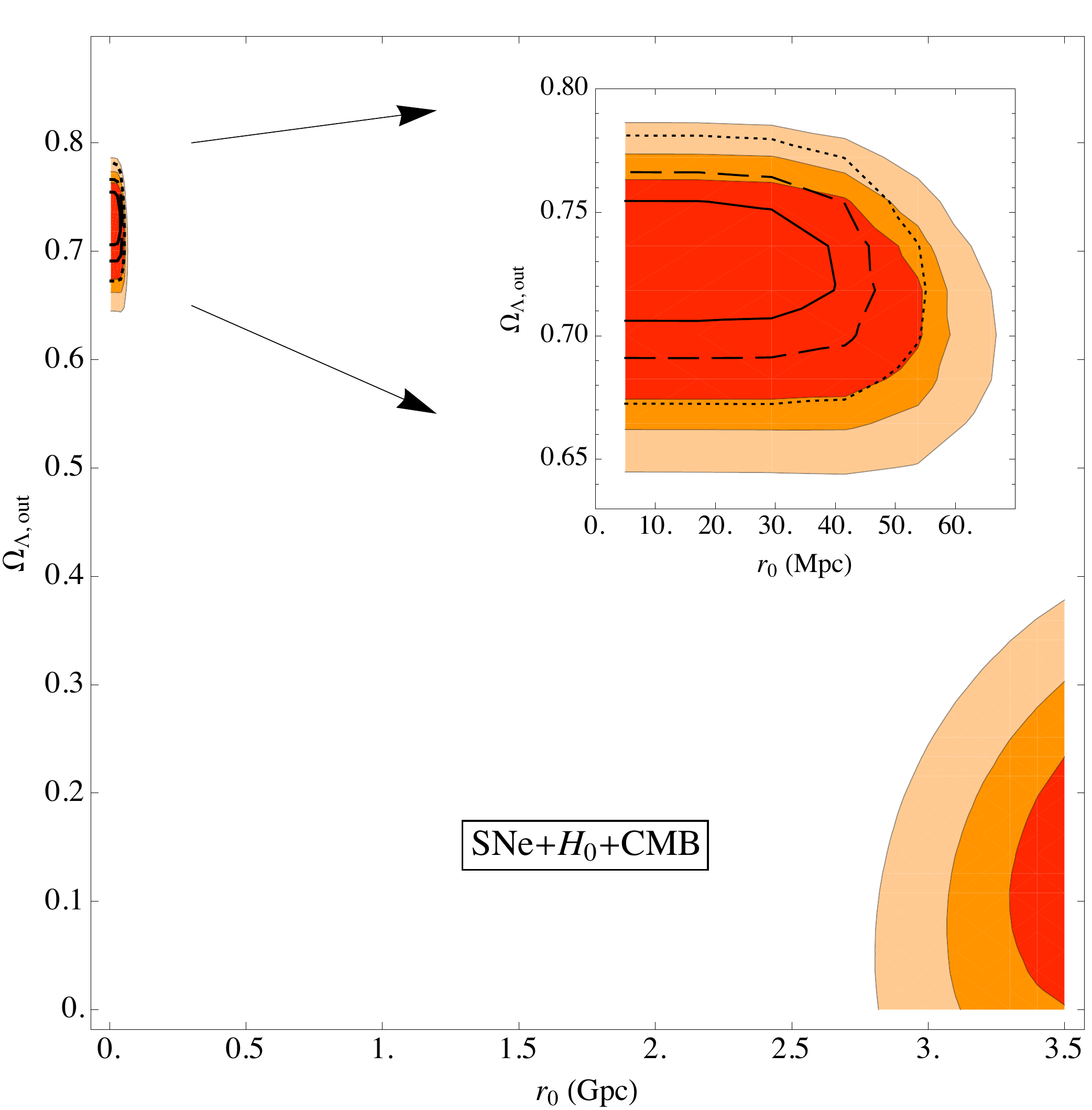}
\includegraphics[width= .49\textwidth]{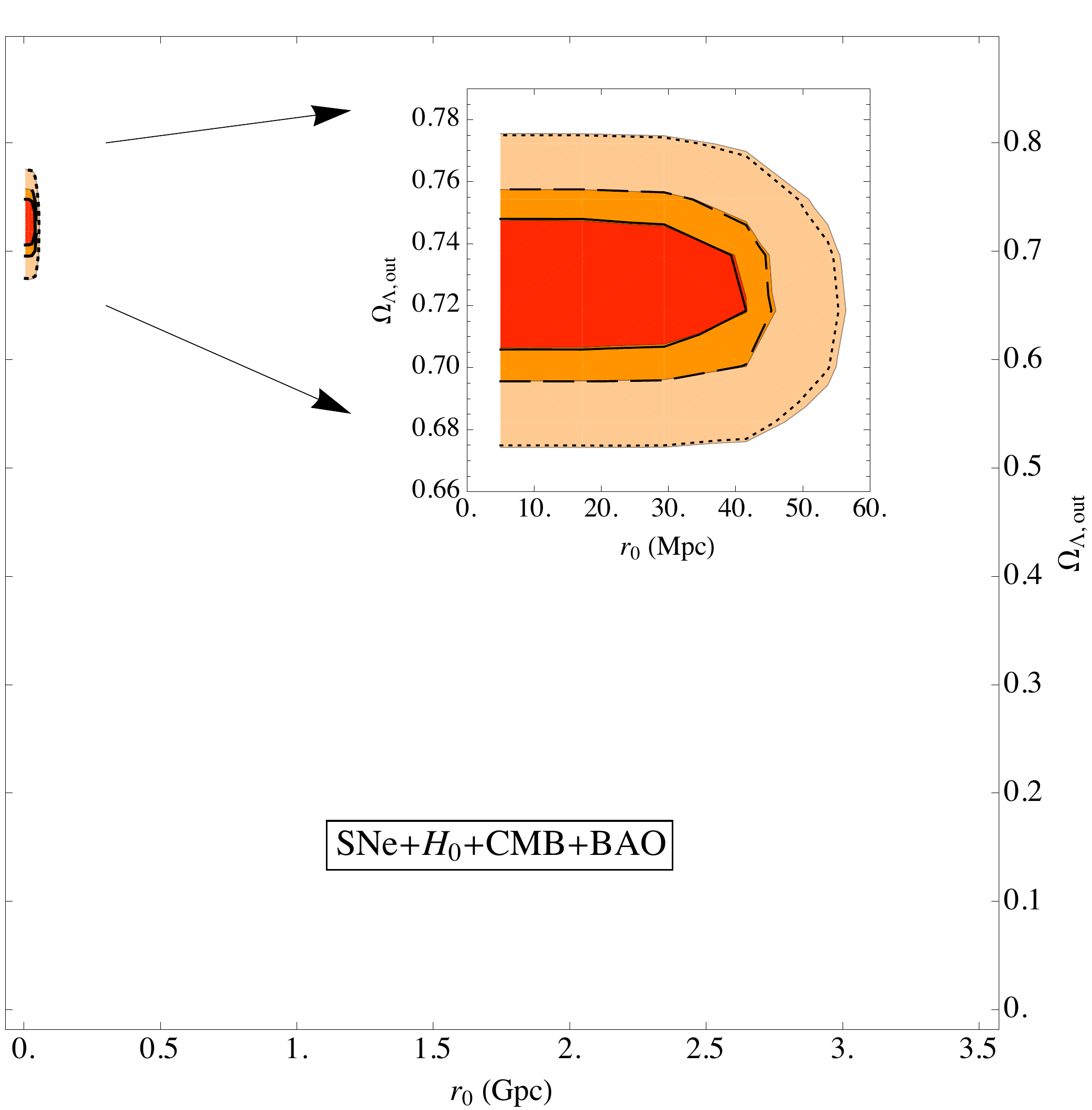}
\caption{
1, 2 and 3$\sigma$ confidence level contours on $r_{0}$ and $\Omega_{\Lambda, \text{out}}$ for the likelihood of Eq.~(\ref{li5}) for the $\Lambda$LTB model described in the text marginalized over $t_{0}$ and $\Omega_{K, \text{out}}$ and with $n_{s}=0.96$. Labelling and definitions as in Fig.~\ref{fig2}.
See Section \ref{nons} for a discussion.
}
\label{fig5}
\end{center}
\end{figure}

Finally, in this Section we will consider models with $n_{s}=0.96$ but of  any age and background curvature, that is, we marginalize the total likelihood over $t_{0}$ and $\Omega_{K, \text{out}}$ as shown in Eq.~(\ref{li5}).
As in the previous Sections, $t_{0} \in [11,\, 16.7]$ Gyr and $\Omega_{K, \text{out}} \in [-0.25,\, 0.16]$.
The results are shown in Fig.~\ref{fig5} where again we include (right panel) and exclude (left panel) BAO constraints in the analysis.

With the BAO constraints the concordance model is favoured at 3$\sigma$ confidence level against a void model of large radius of any type.
The same is true without the BAO constraints if $H_{\text{R09}}$ is used.
In the case of $H_{\text{S06}}$, instead, pure-matter void models are within the 2 and 3$\sigma$ confidence levels, with the 1$\sigma$ confidence level favouring models of large radius and $\Omega_{\Lambda, \text{out}} \approx 0-0.2$.

\section{Conclusions} \label{outlook}

In the present paper we attempted to ``democratically'' confront the concordance $\Lambda$CDM model with the inhomogeneous matter-only alternative represented by the LTB void models.
To this end we performed a likelihood analysis of $\Lambda$LTB models in a seven-dimensional parameter space describing a local void embedded in a possibly curved FLRW background.
We plotted likelihood surfaces with respect to the background cosmological constant $\Omega_{\Lambda, \text{out}}$ and void radius $r_{0}$.
These two parameters best characterize indeed the limiting cases of $\Lambda$CDM and pure-matter void models, which are recovered as delimiting axes of the parameter space. We set the void depth to the value required by a good fit to the SNe observations.

In order to analyze the data with as few priors as possible, we marginalized the total likelihood over the spectral index $n_{s}$, the age of the universe $t_{0}$ and the background curvature $\Omega_{K, \text{out}}$. See Eqs.~(\ref{li1}-\ref{li5}) for a quick summary and legend.
In this way, it is not necessary to compute the reduced $\chi^{2}$, or even its minimum, thus overcoming some of the ambiguities in comparing the two models.
We remind that the LTB model is specified not by free parameters, but by free functions and therefore the choice of the number of degrees of freedom to be adopted in computing the reduced~$\chi^{2}$, which is usually used in model ranking, is necessarily made on a somewhat subjective basis.

We confronted the $\Lambda$LTB model with SNe, Hubble constant, CMB and BAO observations and we found that the concordance model is not the only possibility compatible with the observations. In particular, by allowing a nonzero curvature, we found that a large void with $\Omega_{\Lambda, \text{out}} \approx 0.2-0.4$ lies within the 2$\sigma$ confidence level contour, while a pure-matter model is excluded.

However, these results depend on the precise treatment of the BAO constraints. If we do not include BAO contraints in the analysis -- we remind that perturbation theory in an LTB background is not thoroughly studied yet -- a matter-only model can be as successful as the concordance model.
We stress that our modelling of the void depends crucially only on the void radius and depth, which are the two main physical quantities describing an underdensity, while by a more specified tuning of the void profile it is possible to fit the BAO data without cosmological constant \cite{Biswas:2010xm}.

There is a number of ways one could improve the analysis of the present paper and obtain better results for the void model.
First, one could scan a larger parameter space.
A possibility regarding the void parameters could be to let free the density contrast (see Appendix \ref{degen}) and one regarding the background parameters could be to consider a running spectral index.
Second, one could extend the modeling of the void; we found interesting four directions.
\begin{enumerate}

\item One could drop the simultaneous big bang condition we imposed with Eq.~(\ref{age}).
Simultaneous big bang excludes decaying modes which would be strongly in contradiction with the inflationary paradigm \cite{Zibin:2008vj}, but Gpc scale voids are anyway at odds with the standard scenario.
Without simultaneous big bang one would have a new free function which would likely improve the agreement with the observations.
See for example Ref.~\cite{GarciaBellido:2008nz}.

\item By tuning the density profile one can find better agreement with the BAO observations and with the local measurements of the Hubble rate.
Void models already feature the heavy tuning of the observer position, which has to be close to the center in order not to give a too large dipole. An extra fine tuning on the density profile could be treated in the same pragmatic way.
See for example Ref.~\cite{Biswas:2010xm}.

\item Uncompensated voids can give sizeable redshift effects for observables outside the LTB patch, while this is not generally true for compensated voids. In particular it would be interesting to study very large voids with $r_{0}>3.5$ Gpc. In order to perform such an analysis, however, it would be necessary to generalize the $\Lambda$LTB formalism in order to include the radiation: very large voids extend deeply into the past light cone where radiation cannot be neglected.
See for example Ref.~\cite{Zibin:2008vk}.

\item The possibility that the radiation density is not uniform is also interesting. As in the previous point, it requires a proper modeling of the radiation content.
See for example Ref.~\cite{Clarkson:2010ej}.

\end{enumerate}
By exploiting these extensions it may be possible to successfully fit all the present-day cosmological observables.
In particular, the recent claim \cite{Zhang:2010fa} that void models give a too large kinetic Sunyaev-Zel'dovich effect could be accommodated within a more general modeling.

Finally, we have seen that the areas of the parameter space which give a good fit to the observations are always disconnected.
Our analysis has indeed shown that a local void does not affect the parameter extraction for $\Lambda$CDM models if its radius is smaller than~1-2~Gpc.

Together with this paper we release the Mathematica package \mbox{\tt LLTB 1.0}, which is available at the address \mbox{\tt turbogl.org/LLTB.html}.
We also would like to recommend an interested reader to visit the same webpage, where pre-compiled Mathematica notebooks with animations for exploring a wide range of  the $t_{0}$ and $\Omega_{K, \text{out}}$ parameters are available.

\acknowledgments

The idea at the basis of this work originated at LLTB2009 Workshop held at the KEK (Tsukuba, Japan) to which VM was kindly invited by Hideo Kodama. VM wishes to thank all the participants for the interesting and fruitful discussions.
The authors warmly thank Kimmo Kainulainen for help throughout the development of the manuscript.
The authors benefited from discussions with Alessio Notari, Miguel Quartin, Wessel Valkenburg and James Zibin.

\appendix 
\section{Degeneracy between $\delta_{\Omega}$ and $\Omega_{K, \text{out}}$}
\label{degen} 

In the analysis of Section \ref{results} we set  $\delta_{\Omega}=-0.9$ (which corresponds to a matter contrast of $\delta_{M}\approx -0.81$) because it always guarantees a good fit to the SNe.
If $\Omega_{K, \text{out}} \neq 0$, however, the latter is not a necessary requirement as far as a good fit to the SNe is concerned. The reason is that the background curvature can account for part of the necessary jump in $\Delta H$ demanded to mimic the observed acceleration.

\begin{figure}
\begin{center}
\includegraphics[width= .49\textwidth]{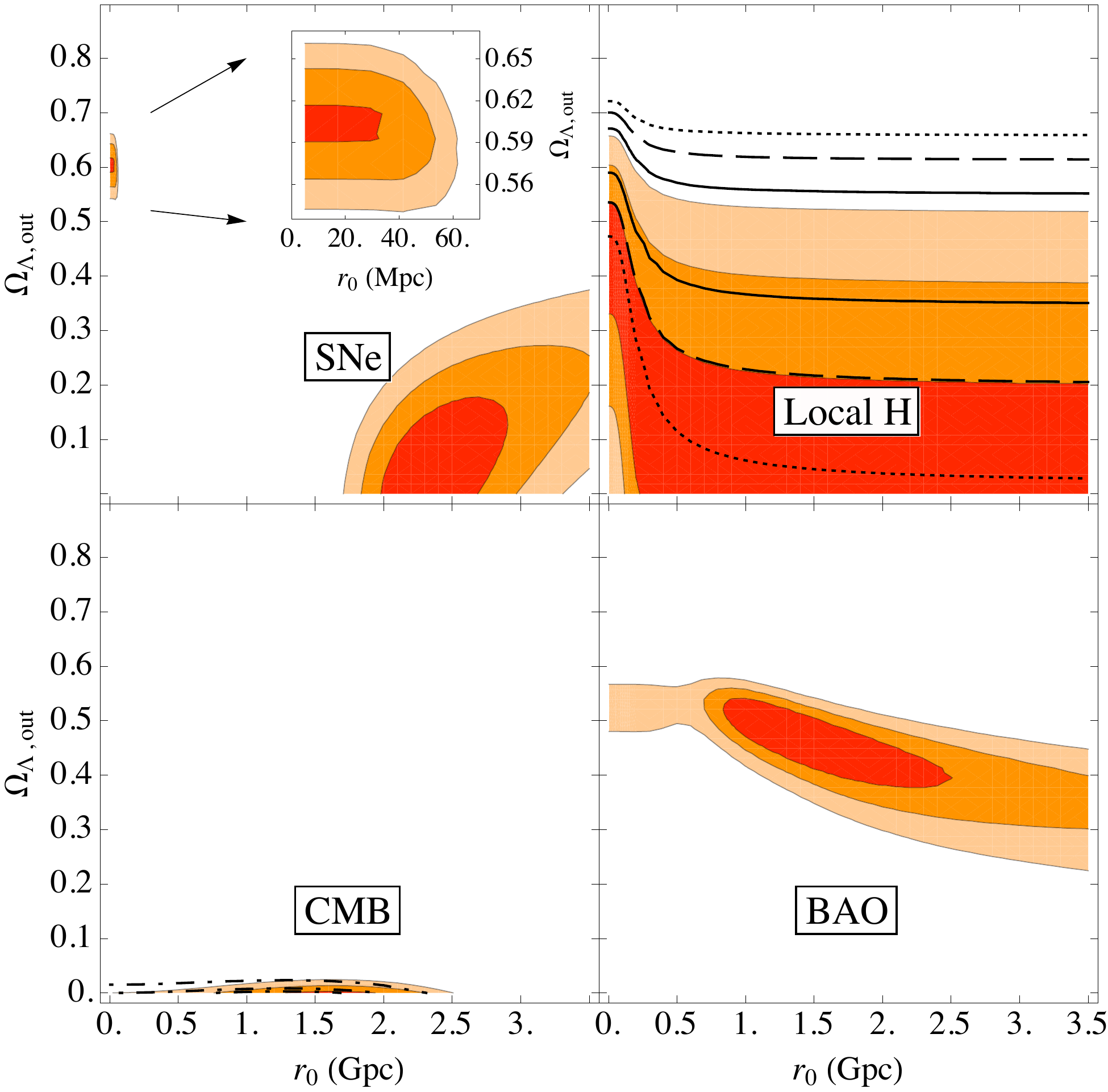}
\includegraphics[width= .49\textwidth]{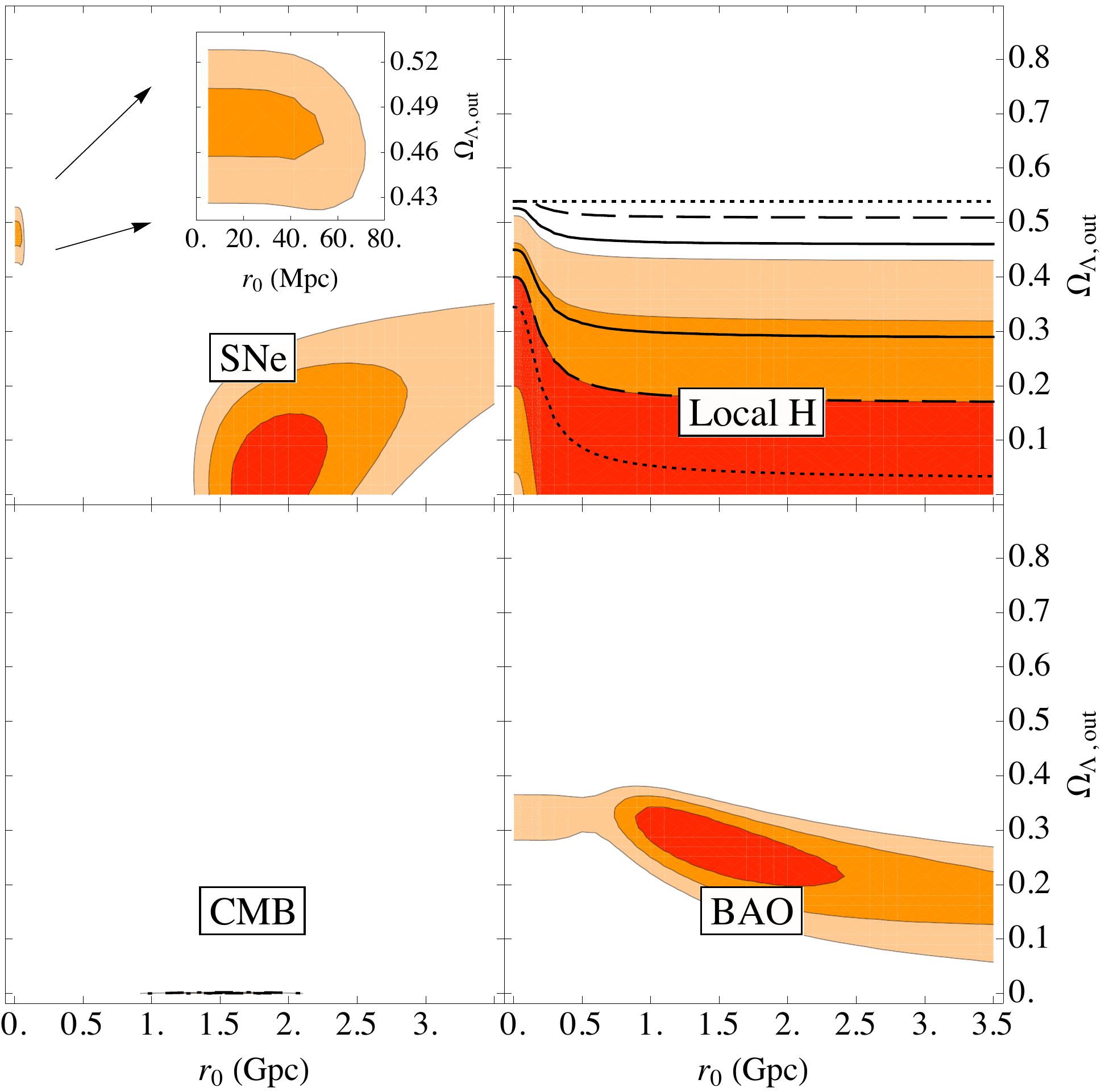}
\caption{
1, 2 and 3$\sigma$ confidence level contours on $r_{0}$ and $\Omega_{\Lambda, \text{out}}$ for the likelihood of Eq.~(\ref{li1}).
For the panels on the left it is $\Omega_{K, \text{out}}=0.2$ and $\delta_{\Omega}=-0.85$ (which corresponds to a matter contrast of $\approx -0.75$).
For the panels on the right it is $\Omega_{K, \text{out}}=0.4$ and $\delta_{\Omega}=-0.8$ (which corresponds to a matter contrast of $\approx -0.7$).
In both cases it is $n_{s}=0.96$ and $t_{0}=13.7$ Gyr.
Labelling and definitions as for the four left panels in Fig.~\ref{fig1}.
See Appendix \ref{degen} for a discussion.
}
\label{fig6}
\end{center}
\end{figure}

In order to explore this possibility we have plotted in Fig.~\ref{fig6} two examples of models with a lower contrast that give good fits to the SNe.
The parameter values are the same of Fig~\ref{fig1} with the difference that, instead of $\Omega_{K, \text{out}}=0$ and $\delta_{\Omega}=-0.9$, it is $\Omega_{K, \text{out}}=0.2$ and $\delta_{\Omega}=-0.85$ (which corresponds to a matter contrast of $\approx -0.75$) for the panels on the left and $\Omega_{K, \text{out}}=0.4$ and $\delta_{\Omega}=-0.8$ (which corresponds to a matter contrast of $\approx -0.7$) for the panels on the right.
As one can see, a lower contrast needs a higher background curvature, thus confirming a degeneracy between $\delta_{\Omega}$ and $\Omega_{K, \text{out}}$.
Moreover, an open background mimics an underdensity and so also the void radius is smaller.

We have seen that an open background allows to have a good SNe fit with a shallower and smaller void, surely a desirable scenario.
To conclude on the viability of this setup we have to look at the other observables.
About local $H$ constraints, by comparing Fig.~$\ref{fig1}$ and Fig.~\ref{fig6} we see that the confidence level contours are basically unaffected as far as pure-matter models are concerned. The reason is that higher $\Omega_{K, \text{out}}$ gives a higher $H_{0, \text{out}}$ which is (al least partially) compensated by the smaller $\Delta H$ of the shallower void.
About the remaining observables, while the confidence level contours relative to the BAO observations change in favor of shallower voids, a good fit to the CMB needs a close background with the result that the models of Fig.~\ref{fig6} are strongly ruled out and the same conclusion holds if we consider different values for the age of the universe.

Concluding, relaxing the prior we have used for the matter contrast could slightly widen the contours towards smaller values of the void radius, but not strongly affect the results of Section \ref{results}.
Finally, we would like to stress that the particular contrast needed for a good fit to the SNe depends on the particular density profile chosen.



\end{document}